\documentclass[aps,
longbibliography,
twocolumn,superscriptaddress,
showpacs]{revtex4-1}
\usepackage{amssymb, graphicx}
\usepackage{enumerate}
\usepackage{amsmath,bm,bbold}
\usepackage{dcolumn}
\usepackage{mathrsfs}
\usepackage{color}
\usepackage[latin1]{inputenc}
\usepackage[normalem]{ulem} 

\usepackage[version=3]{mhchem}
 
\newcommand{\operator}[1]{\ensuremath{\underline{\underline{#1}}}}%
\newcommand{\melement}[3]{\ensuremath{\left\langle #1 \left|#2\right|#3\right\rangle}}%
\newcommand{\ie}{i.\,e.}%

\makeatletter

\newcommand{\ket}[1]{\left|#1\right\rangle}
\newcommand{\bra}[1]{\left\langle #1\right|}
\newcommand{\scalar}[2]{\left\langle #1|#2\right\rangle}

\newcommand{\expected}[1]{\left\langle #1\right\rangle}

\begin{document}

\title{Electron correlation in beryllium: Effects in ground state, short-pulse photoionization and time-delay studies}

\author{Juan J.\ Omiste}
\affiliation{Department of Physics and Astronomy, Aarhus University, 8000  Aarhus C, Denmark}
\affiliation{Present address: Chemical Physics Theory Group, Department of Chemistry,
and Center for Quantum Information and Quantum Control, University of Toronto, Toronto, ON M5S 3H6, Canada}
\author{Wenliang Li}
\affiliation{Department of Physics and Astronomy, Aarhus University, 8000  Aarhus C, Denmark}
\author{Lars Bojer Madsen}
\affiliation{Department of Physics and Astronomy, Aarhus University, 8000  Aarhus C, Denmark}

\date{\today}
\begin{abstract}
We apply a three-dimensional (3D) implementation of the time-dependent restricted-active-space self-consistent-field (TD-RASSCF) method to investigate effects of electron correlation in the ground state of Be as well as in its photoionization dynamics by short XUV pulses, including time-delay in photoionization. First, we obtain the  ground state by propagation in imaginary time. We show that the flexibility of the TD-RASSCF on the choice of the active orbital space makes it possible to consider only relevant active space orbitals, facilitating the convergence to the  ground state compared to the multiconfigurational time-dependent Hartree-Fock method, used as a benchmark to show the accuracy and efficiency of TD-RASSCF. Second, we solve the equations of motion to compute photoelectron spectra of Be after interacting with a short linearly polarized XUV laser pulse. We compare the spectra for different RAS schemes, and in this way we identify the orbital spaces that are relevant for an accurate description of the photoelectron spectra. Finally, we investigate the effects of electron correlation on the magnitude of the relative Eisenbud-Wigner-Smith (EWS) time-delay in the photoionization process into two different ionic channels. One channel, the ground state channel in the ion, is accessible without electron correlation. The other channel is only accessible when including electron correlation.  For theory beyond the mean-field time-dependent Hartree-Fock, the EWS time-delay for the photon energy analyzed is quite insensitive to the considered active orbital spaces.

\end{abstract}
\maketitle
\section{Introduction}
\label{sec:introduction}

The development of short and intense laser pulses has opened the possibility to control and observe electronic and nuclear motion on the attosecond time scale~\cite{Krausz2009,Gallmann2012,Calegari2016}. The new light sources are key to probe and manipulate the electronic structure and dynamics in many-electron atoms~\cite{Shiner2011} and molecules~\cite{Calegari2014,Kraus2015science}. For instance, few photon ionization using short pulses results in ejected electrons whose spatial distribution depends on the energy spectrum and angular momentum of the initial state as well as the ionic channels involved~\cite{Ivanov2013,NgokoDjiokap2015}. Likewise it has become clear that accurate modelling of EWS time-delays in photoionization require careful consideration of electron-correlation effects (see, e.g., Refs.~\cite{Schultze2010,Kheifets2010,Moore2011,Carette2013,Palatchi2014,PhysRevA.94.063409,Ossiander2017} and the recent review Ref.~\cite{Pazourek2015}). Accordingly, 
there is a range of processes that 
require explicit time-dependent methods capable of treating electron correlations beyond mean-field and single-active electron approximations. In the case of  He, it is possible to solve numerically the time-dependent Schr\"odinger equation (TDSE) (see, e.g., Refs.~\cite{Dundas1999,Parker2001a,Feist2008,Peng2011,NgokoDjiokap2011,Pazourek2012,NgokoDjiokap2013a,VanderHart2014}). For larger systems, 
and for photon energies that only affect the valence shell, one can consider two electrons immersed in the mean-field potential produced by the inner electrons~\cite{NgokoDjiokap2013}.  In general, however, in order to consider more than two electrons in an atom and to make the TDSE tractable it is mandatory to use approximations to the wave function and  hence the TDSE. To this end, the time-dependent configuration-interaction (TD-CI) method consists of expanding the many-electron wave function as
\begin{equation}
  \label{eq:wf_ci}
  \ket{\Psi(t)}=\sum_{I\in\mathcal{V}_\textup{FCI}} C_I(t) \ket{\Phi_I},
\end{equation}
where $\mathcal{V}_\textup{FCI}$ denotes the full CI Hilbert space of all accessible configurations. It is the finite size of $\mathcal{V}_\textup{FCI}$ that introduces approximations. In Eq.~\eqref{eq:wf_ci}, $C_I(t)$ is the time-dependent coefficient of the configuration $\ket{\Phi_I}$, formed by a set of spin-orbitals. In this formalism, the TDSE corresponds to a set of first-order differential equations for $C_I(t)$. The description of the continuum, however, demands a large number of orbitals and configurations, which makes the CI approach numerically intractable with increasing number of electrons. To overcome this limitation, there are several methods such as the time-dependent configuration-interaction singles (TD-CIS)~\cite{Pabst2012,Pabst2013}, the time-dependent restricted-active-space configuration-interaction (TD-RAS-CI)~\cite{Hochstuhl2013} and the time-dependent generalized-active-space configuration-interaction (TD-GAS-CI) method~\cite{Bauch2014,Chattopadhyay2015} which impose restrictions on the allowed excitations and the active orbital spaces. These approximations still require many configurations in addition to a careful design of the partitions in the active space. It is also very challenging to extend those methodologies to situations with more than a single electron in the continuum, as is also the case with the $R$-matrix method~\cite{Burke2011} although some progress has been reported in that direction~\cite{VanderHart2014}. Despite the difficulties, it is nevertheless attractive to explore a wave function approach, because of, e.g., the unambiguous extraction of observables. The theory should ideally reduce as much as possible the number of orbitals needed for an accurate description of the configuration space. A breakthrough along those lines came with the multiconfiguration time-dependent Hartree (MCTDH) method~\cite{Meyer1989,Beck1997} and the MCTDH-Fock (MCTDHF) method (see, e.g., Refs.~\cite{Zanghellini2003,Zanghellini2004,Caillat2005,Haxton2011,Haxton2012,Haxton2015}), where time-dependent spin-orbitals are introduced in the ansatz, making $| \Phi_I(t) \rangle$ time-dependent 
\begin{equation}
  \label{eq:wf_mctdhf}
  \ket{\Psi(t)}=\sum_{I\in\mathcal{V}_\textup{FCI}} C_I(t) \ket{\Phi_I(t)}.
\end{equation}
The main advantage of this method is that the use of time-dependent spin-orbitals makes it possible to describe the wave function and, in particular, the continuum with a smaller number of orbitals and configurations than with the time-independent orbitals used in the CI 
approach corresponding to  Eq.~\eqref{eq:wf_ci}. The MCTDHF method has, e.g., been applied to describe high-harmonic generation (HHG) in low dimensions~\cite{Sukiasyan2010}, polarization of the continuum~\cite{Li2016} and to calculate cross sections of atomic~\cite{Hochstuhl2010,Haxton2012} and molecular systems~\cite{Haxton2011}. Since the spatial orbitals are time-dependent, the one- and two-body operators must be updated at each time step, leading to a high numerical cost, especially in the 3D case. Moreover, as in the CI picture, the use of many orbitals may still imply an intractable number of accessible configurations. There are several strategies to diminish the numerical effort without loss of accuracy, such as the time-dependent occupation-restricted multiple-active-space (TD-ORMAS)~\cite{Sato2015}, the time-dependent complete-active-space self-consistent-field (TD-CASSCF)~\cite{Sato2013, Sato2016a} and the time-dependent restricted-active-space self-consistent-field (TD-RASSCF)~\cite{Miyagi2013,Miyagi2014,Miyagi2014b,HaruPRA2017} methods. In particular, the TD-RASSCF method benefits from the RAS to diminish the accessible configurations and hence allows the consideration of only a subset $\mathcal{V}_\text{RAS}$ of the configurations in $\mathcal{V}_\text{FCI}$ by dividing the active space into two or more parts.  Among these parts electron excitations take place with certain restrictions~\cite{Miyagi2013,Miyagi2014,Miyagi2014b,HaruPRA2017}, specified at will, and most often chosen by physical insight into the problem at hand.

In the present work we apply the TD-RASSCF method with double excitations (TD-RASSCF-D) to address the role of electron correlation in Be in the ground state, in photoionization spectra induced by short XUV pulses and in time-delay studies. The TD-RASSCF-D method was previously shown to be accurate and computationally efficient in 1D cases~\cite{Miyagi2013,Miyagi2014b}, and the TD-RASSCF theory was recently extended with a space partitioning concept~\cite{HaruPRA2017}. The main advantage of the approach resides in the possibility of selecting an appropriate RAS which captures the most important configurations for a given system and physical process. For a very recent discussion of these aspects in the case of cold atomic bosons see Ref.~\cite{CamilleNJP2017}. The flexibility in choosing the RAS is remarkable in the imaginary time propagation (ITP), where we show that the active space concept facilitates the convergence. By identifying the most important active orbitals, the TD-RASSCF-D method can be as accurate as the MCTDHF method, but with a smaller number of configurations highlighting the important role of double excitations for the ground state. The application of this method to 3D systems is, however, still computationally challenging due to the large number of non-zero matrix elements of the two-body operator. To overcome this issue, we develop and use the \emph{coupled basis} method to diminish the number of operations required to evaluate the two-body operator. In addition to the ITP studies for the ground state, we present studies of Be subject to short linearly polarized XUV laser pulses. In general the photoelectron energy spectra (PES) reveal detailed properties of atomic, molecular or solid targets, including energies, structural and symmetrical properties of the states of the system. For instance, the directional distribution of the ejected electrons depends on the angular momentum of the final and initial states and the number of absorbed photons~\cite{Hochstuhl2013}. As for the ground state studies, the present TD-RASSCF-D method allows the identification of the most important active orbitals for an accurate description of the photoelectron spectra. Furthermore, the electronic dynamics of the remaining ion during the photoionization has an impact on the outgoing electrons, enclosed, in particular, in the apparent time of ionization~\cite{Schultze2010}. The experimental and theoretical determination of this quantity constitutes a fundamental probe of the many-body dynamics~\cite{Kheifets2010,Moore2011,Nagele2012,Pazourek2015,Nicolaides2010,Komninos2011}. In this work we investigate the relative EWS 
time-delay in ionization between the channel 
(i) Be[$(1s^2 2s^2$)$^1$S$^e$] $\rightarrow$ [Be$^+$($1s^2 2s$) + $\epsilon p$ ] $^1$P$^o$, and the channel 
(ii) Be[$(1s^2 2s^2$)$^1$S$^e$] $\rightarrow$ [Be$^+$($1s^2 2p$) + $\epsilon \ell$ ] $^1$P$^o$ with $ \ell \in (s, d)$.
As is evident from dipole selection rules, process (i) is possible without electron-electron correlation, while in process (ii)  ionization is 
accompanied   by a shakeup in the ion and requires electron-electron correlation.
This study shows that an accurate description of time-delays in Be requires a treatment beyond the mean-field time-dependent Hartree-Fock theory.

The paper is organized as follows. In Sec.~\ref{sec:the_system_and_the_method} we describe the Hamiltonian, the TD-RASSCF-D method and the numerical techniques used, including the coupled basis method. In Sec.~\ref{sec:results} we present the results for Be, both concerning the ground state studies 
(Sec.~\ref{sec:results}.A), the PES (Sec.~\ref{sec:results}.B) and the relative time-delays in ionization (Sec.~\ref{sec:results}.C). Sec.~\ref{sec:conclusions_and_outlook} summarizes the main findings and concludes. Atomic units are used throughout unless indicated otherwise.

\section{Hamiltonian and TD-RASSCF-D}
\label{sec:the_system_and_the_method}

\begin{figure}
  \centering
  \includegraphics[width=0.9\linewidth]{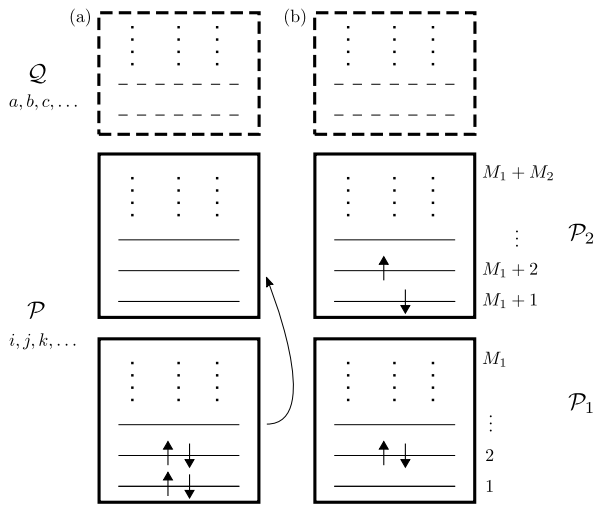}
  \caption{\label{fig:fig1} Restricted-active-space (RAS) associated with the present considerations for the Be atom. The RAS is divided in two partitions: $\mathcal{P}_1$ which is formed by $M_1$ spatial orbitals ($\phi_1,\ldots \phi_{M_1}$) and $\mathcal{P}_2$ by $M_2$ spatial orbitals ($\phi_{M_1+1},\ldots \phi_{M_2+M_1}$). The figure illustrates (a) all the electrons in the lowest-energy configuration and (b) an example of a doubly-excited configuration. The indexes to the left indicate the notation used to label the orbitals in the different spaces. The formalism also allows for a core with always occupied time-dependent orbitals corresponding to a $\mathcal{P}_0$ space~\cite{Miyagi2013,Miyagi2014b}. In this work we do not invoke the $\mathcal{P}_0$ space and therefore $\mathcal{P}_0$ is not shown in the figure.}
\end{figure}

The TD-RASSCF theory for fermions was presented in Refs.~\cite{Miyagi2013, Miyagi2014,Miyagi2014b}, and the similarities and differences between the fermionic and bosonic cases were discussed in Ref.~\cite{CamilleNJP2017},
so the presentation here will be brief. We consider an $N_e$ electron atom with a nuclear charge $Z$ interacting with a linearly polarized laser pulse of short duration. We take the spin-restricted ansatz of the many-electron wave function as
\begin{equation}
  \label{eq:wf}
  \ket{\Psi(t)}=\sum_{I\in\mathcal{V}_\textup{RAS}} C_I(t) \ket{\Phi_I(t)},
\end{equation}
with the configurations $\ket{\Phi_I(t)}$ constructed from $M$ spatial orbitals which can each hold a spin up and a spin down electron, i.e., $2M$ time-dependent spin-orbitals, $\{\ket{\phi_j(t)}\}_{j=1}^{2M}$, 
which form an orthonormal basis in the $\mathcal{P}$ space, with the $\mathcal{Q}$ space being the rest of the single-particle Hilbert space in Fig.~\ref{fig:fig1}. 
The sum in Eq.~\eqref{eq:wf} runs over all accessible configurations specified by the Hilbert space with restrictions on the active space, 
$\mathcal{V}_\text{RAS}$. 

In the MCTDHF method~\cite{Caillat2005,Haxton2011,Haxton2012} the active space, formed by the $M$ time-dependent spatial orbitals, constitutes the full $\mathcal{P}$ space, \ie, the configurations considered in MCTDHF are all the combinations of orbitals allowed by the Pauli principle, as expressed in Eq. (2) by letting the multiindex $I$ run over the full configurational space $\mathcal{V}_\textup{FCI}$.   In contrast, in the present TD-RASSCF-D approach we divide the active space in $\mathcal{P}_1$ and $\mathcal{P}_2$ spaces, which contain $M_1$ and $M_2$ spatial orbitals, respectively (see Fig.~\ref{fig:fig1}).  The sum of the spatial orbitals in the two spaces fulfills $M=M_1+M_2$, and the excitation from $\mathcal{P}_1$ to $\mathcal{P}_2$ is subject to restrictions as specified by the RAS scheme. In the case of the doubles scheme, we allow only double-electron excitations from $\mathcal{P}_1$ to $\mathcal{P}_2$. These restrictions reduce the configurational space and means that the multiindex $I$ in Eq. (3) runs over the RAS, $\mathcal{V}_\textup{RAS}$. In Fig. 1, we may understand the reduction in configurations when changing from the MCTDHF to the TD-RASSCF-D case by noting that in the latter case the configurations with all 4 electrons occupying orbitals in $\mathcal {P}_2$ are absent by restriction on the active orbital space. The RAS concept can be extended by a $\mathcal{P}_0$ space describing a core where the orbitals are always occupied.  In this paper we consider TD-RASSCF-D without a core, that is, we allow double-electron excitations from $\mathcal{P}_1$ to $\mathcal{P}_2$. The TD-RASSCF-D method is numerically efficient and accurately accounts for the two-body interactions. 
 Also note that configurations with double excitations with time-dependent orbitals include non-vanishing projections on the singles space spanned by time-independent orbitals~\cite{Miyagi2013,Miyagi2014b}.

To apply the TD-RASSCF-D method, it is convenient to write the Hamiltonian in second quantization
\begin{equation}
  \label{eq:hamiltonian}
H(t)=\sum_{pq}h_q^p(t)c_p^\dagger c_q+\cfrac{1}{2}\sum_{pqrs}v_{qs}^{pr}(t)c_p^\dagger c_r^\dagger c_s c_q,
\end{equation}
where $c_p$ and $c_p^\dagger$ are fermionic annihilation and creation operators of an electron in the time-dependent spin-orbital $\ket{\phi_p(t)}$. We use the notation $p,q,r,s\ldots$ for all the orbitals and $i,j,k,l\ldots$ for the orbitals in $\mathcal{P}$ space, where we specify by single ($'$) or double prime ($''$) that they belong to different partitions if necessary. In Eq.~\eqref{eq:hamiltonian} the matrix elements are given by
\begin{eqnarray}
\label{eq:h_p_q}
  h_q^p(t)&=&\int\phi_p^\dagger(z,t) h(\vec{r},t)\phi_q(z,t)\mathrm{d}z,\\
\label{eq:v_pr_qs}
\label{eq:mean_field_def}
v_{qs}^{pr}(t)&=&\int \phi_p^\dagger(z_1,t)\phi_q(z_1,t)W_{rs}(z_1,t)\mathrm{d}z_1,
\end{eqnarray}
where $z=(\vec{r},\sigma)$ denotes space and spin degrees of freedom, $W_{rs}(z_1,t)$ is the mean-field operator 
\begin{equation}
\label{meanfield}
W_{rs}(z_1,t) = \int \frac{\phi_r^\dagger(z_2,t) \phi_s(z_2,t)}{|\vec{r}_1-\vec{r}_2|} dz_2
\end{equation}
and  $h(\vec{r},t)$ is the one-body Hamiltonian in the length gauge
\begin{equation}
  \label{eq:h_one_body}
  h(\vec{r},t)=-\cfrac{1}{2}\nabla^2-\cfrac{Z}{r}+\vec{E}(t)\cdot\vec{r},
\end{equation}
with $Z$ the nuclear charge, $r= | \vec{r} |$ the magnitude of the position vector, and $\vec{E}(t)$ the electric field of the laser pulse. 

We now briefly describe the derivation of the equations of motion. The  TD variational principle~\cite{Dirac1930,Raab2000} establishes that the best approximation to the time-dependent Schr\"odinger equation (TDSE) is a stationary point of the functional
  \begin{eqnarray*}
 \mathcal{S}[\{C_I\},\{\phi_i\},\{\epsilon^i_j\}]&=&
\int_0^T  \left[ \langle\Psi(t)|\left(i\cfrac{\partial}{\partial t}-H\right)|\Psi(t)\rangle \right.
\\
&+& \left. \sum_{ij}\epsilon_j^i(t)\left( \langle\phi_i(t)|\phi_j(t)\rangle-\delta_{ij}     \right) \right]\mathrm{d}t,
  \end{eqnarray*}
where the Lagrange multipliers $\epsilon_{j}^i(t)$ ensure that the orbitals in the $\mathcal{P}$ space are orthonormal during the time interval $[0,T]$. To simplify the notation, we do not indicate the time, space or spin dependence of the amplitudes and the orbitals. Then, we seek stationary points, $\delta \mathcal{S}=0$, and in the TD-RASSCF-D case the stationary points fulfil the equations~\cite{Miyagi2013,Miyagi2014b}
\begin{eqnarray}
\label{eq:ci_dot}
i\dot C_I &=&\sum_{ij}(h^i_j - i\eta_j^i)\langle\Phi_I |  c_i^\dagger c_j  | \Psi\rangle \nonumber \\ 
&+&\cfrac{1}{2}\sum_{ijkl}v_{jl}^{ik}\langle\Phi_I|  c_i^\dagger c_k^\dagger c_l c_j |\Psi\rangle,
\end{eqnarray}
\begin{equation}
     \label{eq:q_space}
     i\sum_j Q |{\dot\phi_j}\rangle\rho_i^j=\sum_jQh(t)|{\phi_j}\rangle\rho_i^j+\sum_{jkl}QW_l^k|{\phi_j}\rangle\rho_{ik}^{jl},
\end{equation}
\begin{equation}
\label{eq:p_space}
\sum_{k''l'}(h_{l'}^{k''}-i\eta_{l'}^{k''})A_{k''i'}^{l'j''}+\sum_{klm}(v_{kl}^{j''m}\rho_{i'm}^{kl}-v_{i'm}^{kl}\rho_{kl}^{j''m})=0, 
\end{equation}
with $\eta_j^{i}=\langle \phi_i|\dot{\phi}_j\rangle$, $Q=\mathbb{1}-P=\mathbb{1}-\sum_{j} \ket{\phi_j}\bra{\phi_j}$
$\rho_i^j=\langle\Psi|c_i^\dagger c_j|\Psi\rangle$, $\rho_{ik}^{jl}=\langle\Psi|c_i^\dagger c_k^\dagger c_l c_j|\Psi\rangle$ and
$A_{ki}^{lj}= \langle\Psi|[c_i^\dagger c_j, c_k^\dagger c_l]|\Psi\rangle$.
The amplitude equation ~\eqref{eq:ci_dot} describes the time evolution of the expansion coefficients $C_I(t)$ of Eq.~\eqref{eq:wf}, and the orbital equations~\eqref{eq:q_space}-\eqref{eq:p_space} describe the projection of the time derivative of the orbitals on the $\mathcal{Q}$ and $\mathcal{P}$ spaces, respectively. 
The Eqs.~\eqref{eq:ci_dot}-\eqref{eq:p_space} constitute a set of coupled non-linear differential equations and the strategy to solve them is the following: i) first, we use Eq.~\eqref{eq:p_space} to solve for $\eta_{l'}^{k''}$, ii) then, we can solve the amplitudes~\eqref{eq:ci_dot} and the $\mathcal{Q}$ space Eqs.~\eqref{eq:q_space} (see the discussion in Ref.~\cite{CamilleNJP2017}). On the one hand, let us consider the case of MCTDHF, that is, if we take into account all the combinations of the orbitals to construct the Slater determinants, then Eq.~\eqref{eq:p_space} becomes an identity~\cite{Beck2000a,Miyagi2013,Miyagi2014b}. Therefore, in this case we only need to solve Eqs.~\eqref{eq:ci_dot} and~\eqref{eq:q_space}, and we can set $\eta_{j}^i$ to be any antihermitian matrix~\cite{Miyagi2013} and in practice often $\eta_j^i=0$ is used. On the other hand, for the TD-RASSCF-D method including only double excitations~\cite{Miyagi2013,Miyagi2014b}, the constraints on the active space imply that we only have the freedom to choose a fixed value for $\eta_{i}^j$ if both the orbitals $i$ and $j$ belong to either $\mathcal{P}_1$ or $\mathcal{P}_2$. The calculation of $\eta_{i'}^{j''}$ for $i'$ and $j''$ belonging to different partitions implies solving Eq.~\eqref{eq:p_space}, together with Eqs.~\eqref{eq:ci_dot} and~\eqref{eq:q_space}. 

The main advantage of the TD-RASSCF-D method compared with MCTDHF is that it is possible to include only the relevant configurations for a given physical observable or process. The numerical effort is therefore smaller for the same number of orbitals~\cite{Miyagi2013,Miyagi2014b}, a point we will discuss further below. This method shares with the MCTDHF the numerical instability in the $\mathcal{Q}$ space equation induced by the singularity of the inverse of the reduced one-body density matrix, $({\operator{\rho}})^{-1}$. In particular, a system with low entanglement would require only a small number of orbitals, and the addition of more leads to small eigenvalues of the one-body density matrix~\cite{Hinz2016}. To address this problem, the one-body density matrix is usually regularized $\operator{\widehat{\rho}}=\operator{\rho}+\epsilon e^{-\operator{\rho}/\epsilon}$, where $\epsilon$ is a regularization parameter~\cite{Beck2000a}. In this work we set $\epsilon=10^{-10}$.

\subsection{Single-orbital basis}
\label{sec:single_orbital_basis}
Here we discuss the single-orbital basis and for completeness we collect the formulas needed for its construction. To describe the orbitals of the many-electron wave function, we use a basis set expansion in a finite element discrete variable representation (FE-DVR) for the radial grid~\cite{McCurdy2004} and spherical harmonics for the angular part,~\ie, 
\begin{equation}
  \label{eq:phip}
\scalar{\vec{r}}{\phi_j(t)}=  \phi_j(\vec{r},t)=\sum_{\alpha \ell m} c^{j}_{\alpha \ell m}(t)\cfrac{\chi_\alpha(r)}{r}Y_{\ell m}(\Omega),
\end{equation}
where $c^{j}_{\alpha \ell m}(t)$ are the coefficients of the expansion, $\chi_\alpha(r)$ are the FE-DVR functions and $Y_{\ell m}(\Omega)$ are the spherical harmonics. To define the FE-DVR functions we divide the radial grid $[0,r_\text{max}]$ in $N_{fe}$ finite elements with $N_b$ nodes in each element. Then, the collective subscript $\alpha=(a,e)$ in Eq.~\eqref{eq:phip} denotes the node $a=1,\ldots, N_b$ in the element $e=1,\ldots,N_e$. The FE-DVR function is defined using Gauss-Lobatto functions~\cite{McCurdy2004}
\begin{equation}
\chi_{a,e}(r) =
                  \left\{\begin{array}{l}
                    \cfrac{1}{\sqrt{\omega_{a,e}}}\prod_{j\ne a}\cfrac{r-r_{j,e}}{r_{a,e}-r_{j,e}}, r_{1,e}\le r\le r_{N_b,e}\\
\\
0,~\text{otherwise}
                  \end{array}\right.
\end{equation}
where $r_{a,e}$ are the nodes and $\omega_{a,e}$ the weights of the associated Gauss-Lobatto quadrature. The FE-DVR functions have the following properties~\cite{Rescigno2000}
\begin{eqnarray}
  \label{eq:fedvr_value}
  &&  \chi_{a,e}(r_{b,e'}) = \cfrac{\delta_{a,b}\delta_{e,e'}}{\sqrt{\omega_{a,e}}},\\
\label{eq:fedvr_orthonormalization}
&&\int \mathrm{d}r \chi_{a,e}(r)\chi_{b,e'}(r) = \delta_{a,b}\delta_{e,e'},\\
\label{eq:fedvr_matrix_element}
&& \int \mathrm{d}r \chi_{a,e}(r)\chi_{b,e'}(r) f(r)= f(r_a)\delta_{a,b}\delta_{e,e'}.
\end{eqnarray}
We obtain the radial part of the kinetic energy term by calculating the matrix element of the second derivative in the radial coordinate~\cite{Rescigno2000}
\begin{eqnarray}
  \nonumber
&&  -\cfrac{1}{2}\int \mathrm{d}r \chi_{a,e}(r)\cfrac{\partial^2}{\partial r^2}\chi_{b,e'}(r)=\\
&&\cfrac{1}{2}(\delta_{e,e'}+\delta_{e,e'\pm 1})\int\mathrm{d}r \cfrac{\partial}{\partial r}\chi_{a,e}(r)\cfrac{\partial}{\partial r}\chi_{b,e'}(r),
\end{eqnarray}
where~\cite{Rescigno2000}
\begin{eqnarray}
\nonumber
&&  \cfrac{\partial}{\partial r}\chi_{a,e}(r_{a',e'})=\\
  \label{eq:fedvr_derivative}
&&\left\{
  \begin{array}{l}
    \cfrac{1}{\sqrt{\omega_{a,e}}(r_{a,e}-r_{a',e'})}\prod_{k\ne a,a'}\cfrac{r_{a',e}-r_{k,e}}{r_{a,e}-r_{k,e}}, a\ne a'\\
\cfrac{1}{2\omega_{a,e}^{3/2}}(\delta_{a,N_b}-\delta_{a,1}), a=a'
  \end{array}
\right.
\end{eqnarray}
We introduce the bridge functions~\cite{McCurdy2004}
\begin{equation}
  \label{eq:bridge_functions}
  \tilde\chi_{N_b,e}(r)=\tilde\chi_{1,e+1}(r)=\cfrac{\sqrt{\omega_{N_b,e}}\chi_{N_b,e}(r)+\sqrt{\omega_{1,e+1}}\chi_{1,e+1}(r)}{\sqrt{\omega_{N_b,e}+\omega_{1,e+1}}},
\end{equation}
which ensure the continuity of the functions between adjacent elements, and fulfil Eqs.~\eqref{eq:fedvr_value}-\eqref{eq:fedvr_matrix_element}.

\subsection{Evaluation of the mean-field operator}
\label{sec:mean_field_operator}
The matrix elements of the two-body operator need to be updated at each time step and this update is the main bottleneck of time-dependent SCF methods. The two-body electron-electron interaction reads in the multipole expansion
\begin{equation}
  \label{eq:multipole_expansion}
  \cfrac{1}{|\vec{r}-\vec{r'}|}=\sum_{L=0}^\infty\cfrac{4\pi}{2L+1}\cfrac{r_<^L}{r_>^{L+1}} \sum_{M=-L}^L Y_{LM}(\Omega)Y_{LM}^\dagger(\Omega'),
\end{equation}
where $r_>$ ($r_<$) is the largest (smallest) between $|\vec{r}|~\text{and}~|\vec{r'}|$. The electron-electron Coulomb repulsion is not diagonal in the angular coordinates, because, in the evaluation of the matrix elements, each spherical harmonic in Eq.~\eqref{eq:multipole_expansion} couples with two angular functions coming from the product of two orbitals. This leads to numerous non-vanishing matrix elements which require both large memory for storage and long CPU time for updating the orbitals. There are several proposals to overcome these issues. On the one hand, we can reduce the number of applications of the two-body operator using the constant mean-field approximation~\cite{Beck2000a} or by restricting the number of operations by allowing only a small number of $m$ in the single-orbital expansion~\cite{Sato2016a}, or both. On the other hand, we can make the two-body operator sparse by expanding it in a pseudo-DVR basis~\cite{Haxton2007,Hochstuhl2014} for the angular grid. 
In this work, we apply what we call the \emph{coupled basis} method. We use that the electron-electron interaction commutes with the total angular momentum and its projection on the $z$ axis for two electrons,~\ie, $\left[(\vec{l}_1+\vec{l}_2)^2,1 /|\vec{r_1}-\vec{r_2}|\right]=\left[(l_{z,1}+l_{z,2}),1/|\vec{r_1}-\vec{r_2}|\right]=0$. This implies that $1/|\vec{r_1}-\vec{r_2}| $ conserves the coupled $\ell~\text{and}~m$ in a basis of two-orbital functions in the 
coupled representation. At each time step, we therefore transform the product of two orbitals into a two-orbital function in the coupled representation
\begin{eqnarray}
\nonumber
&& \phi_i(\vec{r},t)^\dagger\phi_j(\vec{r},t)|_{r=r_\alpha}=\\
\nonumber
&&=\sum_{\alpha\ell \ell' mm'}c_{\alpha\ell m}^{i\dagger}c_{\alpha\ell'm'}^{j}\cfrac{\chi_\alpha(r_\alpha)^2}{r_\alpha^2}Y_{\ell m}(\Omega)^\dagger Y_{\ell'm'}(\Omega)\\
\label{eq:pq_coupled}
&&=\sum_{\alpha}\sum_{L'=0}^{2\ell_\text{max}}\sum_{M'=-L'}^{L'}\cfrac{\omega_\alpha^{-1}}{r_\alpha^2}\Theta^{ij}_{\alpha L'M'}Y_{L'M'}(\Omega'),
 \end{eqnarray}
where we used the FE-DVR property that $\chi_\alpha(r_\gamma)\chi_\beta(r_\delta)=\delta_{\alpha,\beta}\delta_{\gamma,\delta}\delta_{\alpha,\gamma}\chi_\alpha(r_\alpha)^2=\delta_{\alpha,\beta}\delta_{\gamma,\delta}\delta_{\alpha,\gamma}\omega_\alpha^{-1}$, and where we defined
\begin{equation}
\Theta^{ij}_{\alpha L'M'}=\sum_{\ell,m,\ell',m'}c_{\alpha\ell m}^{i\dagger} c^{j}_{\alpha\ell'm'}(-1)^{M'} y_{\ell m,\ell'm',L'-M'},
\end{equation}
with $y_{\ell m,\ell'm',L'M'}$ the Gaunt coefficients
\begin{eqnarray}
\nonumber
&&  y_{\ell_1m_1,\ell_2m_2,\ell_3m_3}=\int \mathrm{d}\Omega Y_{\ell_1m_1}^\dagger(\Omega)Y_{\ell_2m_2}(\Omega)Y_{\ell_3m_3}(\Omega).
  \label{eq:gaunt_coefficients}
\end{eqnarray}
From Eq.~\eqref{eq:multipole_expansion} we now see that the two-body matrix elements in the coupled basis form a  block diagonal matrix in $(LM)$. Using this property, the mean-field operator in Eq.~\eqref{meanfield} reads as
\begin{eqnarray}
\nonumber
W_{ij}(\vec{r})&=&\int\mathrm{d}\sigma W_{ij}(z)=\int\mathrm{d}^3\vec{r'}\cfrac{\phi_i(\vec{r'})^\dagger\phi_j(\vec{r'})}{|r-r'|}=\\
\label{eq:mfo_expanded_fedvr}
&=&  \sum_{\alpha}\sum_{L=0}^{2\ell_\text{max}}\sum_{M=-L}^L\Theta^{ij}_{\alpha L'M'}f_{\alpha,L}(r)Y_{LM}(\Omega),
\end{eqnarray}
where $f_{\alpha,L}(r)=\int_{r'}\chi_\alpha(r')^2\frac{4\pi}{2L+1}\frac{r_<^L}{r_>^{L+1}}dr'$.  It is useful to rewrite Eq.~\eqref{eq:mfo_expanded_fedvr} in the FE-DVR basis. After some algebra we find
\begin{equation}
\label{eq:mfo_coupled_fedvr_omega}
W_{ij}(\vec{r})\approx\sum_{\gamma,L'',M''}\cfrac{\chi_\gamma(r)^2}{r^2}Y_{L''M''}(\Omega)\bar\omega^{ij}_{\gamma L'' M''},
\end{equation}
with
\begin{eqnarray}
\label{eq:omega_pq_lm}
\bar\omega^{ij}_{\gamma L M}&=&\sum_{\alpha}\Theta^{ij}_{\alpha LM}R_{L}(\alpha,\gamma), \\
\label{eq:rl_a_gamma}
R_{L}(\alpha,\gamma)&=&\int\chi_\gamma(r)^2 f_{\alpha,L}(r)\mathrm{d}r,
\end{eqnarray}
where $R_{L}(\alpha,\gamma)$ is approximated by~\cite{McCurdy2004}
\begin{equation}
  \label{eq:rl_a_gamma_expression}
  R_L(\alpha,\gamma)= \left(\cfrac{(2L+1)}{r_\alpha r_\gamma\sqrt{\omega_\alpha \omega_\gamma}}\left[T_{\alpha,\gamma}^{(L)}\right]^{-1}+\cfrac{r_\alpha^L r_\gamma^L}{r_\text{max}^{2L+1}}\right),
\end{equation}
where $T_{\alpha,\gamma}^{(L)}$ is twice the kinetic energy matrix. Now, we can obtain the terms needed to solve Eqs.~\eqref{eq:q_space} and~\eqref{eq:p_space}, such as the application of the mean-field operator on a single spatial orbital
\begin{eqnarray}
\nonumber
  W_{ij}(\vec{r})\phi_s(\vec{r})&=&\sum_\alpha\sum_{\ell m}\cfrac{\chi_\alpha(r)}{r} Y_{\ell m}(\Omega)\\
\label{eq:w_pq_phi_s}
&&\times\sum_{LM\ell'm'} \bar\omega^{ij}_{LM}c^s_{\alpha \ell'm'}\cfrac{y_{\ell m,\ell'm',LM}}{\omega_\alpha}.
\end{eqnarray}
Similarly we obtain for the two-body operator
\begin{equation}
  v_{kj}^{li} = \int \mathrm{d}\vec{r}\phi_l^\dagger(\vec{r})W_{ij}(\vec{r})\phi_k(\vec{r}) 
\approx \sum_{\alpha LM}\cfrac{\Theta^{\alpha l\dagger}_{\alpha LM}\bar\omega^{ij}_{\alpha LM}}{\omega_\alpha}.\label{eq:v_rp_sq}
\end{equation}
To highlight the benefits from using the coupled basis method, we compute the number of operations performed at each update. For simplicity, we assume that $\ell_\text{max}=m_\text{max}$ and take into account all possible combinations of spherical harmonics, even if the associated Clebsch-Gordan coefficient is zero. First, the number of operations required for the transformation to the coupled basis is $\mathcal{O}(M^2n_r n_\theta^2(4n_\theta+1-4\sqrt{n_\theta}))$, where $n_r$ is the number of radial functions, $n_\theta$ and $(4n_\theta+1-4\sqrt{n_\theta})$ are the numbers of angular functions in the single-orbital and coupled basis, respectively. Second, the evaluation of the mean-field operator needs $\mathcal{O}(M^2n_r^2(4n_\theta+1-4\sqrt{n_\theta}))$ operations, and the cost of its application to each orbital requires $\mathcal{O}(M^3n_rn_\theta^2(4n_\theta+1-4\sqrt{n_\theta}))$ operations. Finally, the calculation of the matrix element of the two-body operator requires $\mathcal{O}(M^4n_r(4n_\theta+1-4\sqrt{n_\theta}))$ operations. 
The most important achievement of this method is that the number of operations of the mean-field operator scales linearly with the number of angular functions, contrary to the numerical effort required if we use directly the product basis, where the scaling goes as $\mathcal{O}(M^4 n_r^2n_\theta^{4})$.

\subsection{Removing the stiffness}
\label{sec:removing_the_stiffness}

To describe the electronic structure accurately we require a dense grid in the radial coordinate around the nucleus, which in  momentum space implies highly oscillating functions with large momenta. These functions introduce rapid oscillations in time which require a very small time step to be resolved and this stiffness leads to instabilities in Eqs.~\eqref{eq:q_space}-\eqref{eq:p_space}~\cite{Hochstuhl2014,Miyagi2015}. In many cases, however, these high-energy states do not contribute to the dynamics of interests and they can therefore be removed by a suitable projection. The calculation of the eigenstates of the full Hamiltonian is not possible in the present framework, and even worse, in the presence of the laser, the Hamiltonian would have to be diagonalized at each time step. For these reasons, we consider only the high energy states of the one-body Hamiltonian, which is often a good approximation~\cite{Hinz2013}. Taking this consideration into account, we apply the energy subspace projection described in Refs.~\cite{Hinz2013,Hochstuhl2014}. First, we diagonalize the field-free one-body Hamiltonian to obtain the high-energy one-body states $\ket{\psi_j}$. Second, we define the projector $\mathbb{P}=\mathbb{1}-\sum_j\ket{\psi_j}\bra{\psi_j}$ to remove the contributions with eigenenergies $E_j>E_\text{cutoff}$ from the time-dependent many-body Hamiltonian, $H$, obtaining the stiffness-free Hamiltonian  $H'=\mathbb{P}H\mathbb{P}$. The many-body nature of the problem makes it difficult to apply $\mathbb{P}$ on the Hamiltonian, therefore, we perform the projection on the wave function $\melement{\Psi}{\mathbb{P}H\mathbb{P}}{\Psi}=  \melement{\mathbb{P}\Psi}{H}{\mathbb{P}\Psi}$, which consists of projection out the high energy contribution from each orbital. The energy cutoff is set between 350-1500 a.u..

\subsection{Extraction of the photoelectron spectrum}
\label{sec:extraction_photoelectron_spectrum}
In this section we describe a procedure to obtain the photoelectron spectrum. 
Often used methods consists of projecting on scattering waves, Coulomb waves or plane waves~\cite{Gozem2015,Madsen2007,Argenti2013}. While the projection on scattering states can be applied immediately after the end of the pulse, the projection on Coulomb or plane waves requires the propagation over some periods after the laser is switched off to allow the ejected electron to arrive at the outer region, where the interaction of the outgoing electron with the atom can either be neglected or approximated by the Coulombic monopole term. In this work, we proceed as follows. First, we integrate the one-body density over $N_e-1$ electrons and second, we project on Coulomb or plane waves in the outer region, as specified below, in the remaining coordinate. The photoelectron momentum distribution (PMD) reads
  \begin{eqnarray}
    \label{eq:photoelectron_3d}
   \cfrac{\mathrm{d}^3P}{\mathrm{d}^3\vec{k}}&=&\sum_{ij}\rho_i^j \widetilde{\phi}_i^\dagger(\vec{k},t)\widetilde{\phi}_j(\vec{k},t),
  \end{eqnarray}
where $\rho_i^j$ is defined just after Eq.~(\ref{eq:p_space}) and $\widetilde{\phi}_i(\vec{k},t)$ is defined as
\begin{eqnarray}
  \label{eq:proj_outgoing_scattering}
&&  \widetilde{\phi}_j(\vec{k},t)=\\
\nonumber
&&\int\mathrm{d}^3r \psi_{\vec{k}}(\vec{r})^\dagger \phi_j(\vec{r},t)\Xi(r,r_\text{out},r_\text{max},\Delta),
\end{eqnarray}
where $\psi_{\vec{k}}(\vec{r})$ is an outgoing scattering wave function and $\Xi(r,r_\text{out},r_\text{max},\Delta)$ is a window function introduced to remove boundary effects related to the outer region ($r\ge r_\text{out}$) and the end of the box ($r_\text{max}$)~\cite{Parker2001a}
\begin{eqnarray*}
\nonumber
  &&\Xi(r,r_\text{out},r_\text{max},\Delta)=\\
&&\left\{
  \begin{array}{ll}
    0,&r \le r_\text{out},\\
    1-\cos\left(\cfrac{\pi}{2}\cfrac{r-r_\text{out}}{\Delta}\right),&r_\text{out}\le r \le r_\text{out}+\Delta,\\
    1,&r_\text{out}+\Delta\le r \le r_\text{max}-\Delta,\\
        \cos\left(\cfrac{\pi}{2}\cfrac{r-(r_\text{max}-\Delta)}{\Delta}\right),&r_\text{max}-\Delta\le r \le r_\text{max}.\\
  \end{array}
\right.
\label{eq:hamming_function}
\end{eqnarray*}
In the present work we use $r_\text{out}=20$ and $\Delta=20$. In the case of plane waves $\psi_{\vec{k}}^{PW}(\vec{r})=(2\pi)^{-3/2}e^{i\vec{k} \cdot \vec{r}}$, we use the expansion in terms of the spherical Bessel functions, $j_L(kr)$
\begin{equation}
  \label{eq:plane_wave_expansion}
  e^{i\vec{k} \cdot  \vec{r}}=4\pi\sum_{L,M} i^LY_{LM}(\Omega_r)Y_{LM}^\dagger(\Omega_k)j_L(kr).
\end{equation}
to perform the integral in Eq.~\eqref{eq:proj_outgoing_scattering}. In the case of Coulomb wave functions $\psi_{\vec{k}}^{C}(\vec{r})$~\cite{Madsen2007}, we have
\begin{equation}
  \label{eq:coulomb_wave_expansion}
\psi_{\vec{k}}^{C}(\vec{r})=\sqrt{\cfrac{2}{\pi}}  \sum_{L,M} i^L e^{-i\sigma_L(\eta)}Y_{LM}(\Omega_r)Y_{LM}^\dagger(\Omega_k)\cfrac{F_L(kr,\eta)}{kr}
\end{equation}
where $F_L$ is the regular Coulomb function, $\sigma_L(\eta)=\arg[\Gamma(L+1+i\eta)]$ is the Coulomb phase shift and $\eta=-1/k$. The photoelectron spectrum is obtained by integration over the angular coordinates in momentum space, $\Omega_k$
\begin{equation}
  \label{eq:photoelectron_1d}
    \cfrac{\mathrm{d}P}{\mathrm{d}E}= \int_{\Omega_k}\mathrm{d}\Omega_k k \cfrac{\mathrm{d}^3P}{\mathrm{d}^3\vec{k}},
\end{equation}
and the triply differential energy and angular resolved probability is obtained as
\begin{equation}
  \label{eq:triply_diff_energy}
  \cfrac{\mathrm{d^3}P}{\mathrm{d}E\mathrm{d}\Omega_k}=k\cfrac{\mathrm{d^3}P}{\mathrm{d^3}\vec{k}}.
\end{equation}

The validity of the projection on plane waves to describe the spectrum relies on assuming that the wave packet is far from the atom and that it is not affected by the atomic potential~\cite{Madsen2007,Argenti2013}. This assumption is valid for the laser parameters and the range of the photoelectron spectra analyzed in this work, as we validated by comparison with the results obtained by projection on Coulomb waves.

\section{Results}
\label{sec:results}

In this section, we describe the impact of electron correlation on the ground state of Be and the photoionization process, including time-delays, due to the interaction with a linearly polarized XUV laser pulse of short duration. We use Be for these illustrative calculations because it allows for converged MCTDHF reference data. In this way we can identify the most important part of the orbital space for a given physical observable. We describe the radial coordinate from $r=0$ to $8$ using 8 FEs of length $1$. From $r=8$ to the end of the box we add elements of length $4$, with $N_b=8$. To calculate the ground state we perform an imaginary-time propagation (ITP) of an initial guess function in a box from $r=0~\text{to}~28$. The dynamics due to the interaction with the laser is subsequently described performing a real-time propagation (RTP), where we set the end of the box to $r_\text{max}=200$ by adding 43 elements of length 4 with 8 nodes in each. For the ITP, the angular part of the orbitals is described with a maximum orbital angular momentum $\ell_\text{max}=2$ and magnetic quantum numbers $|m_\text{max}|=1$ for 7 or less orbitals and $\ell_\text{max}=3$ and $|m_\text{max}|=2$ otherwise. For the RTP, we use $\ell_\text{max}=3$ and $|m_\text{max}|=2$ in all the cases, and this is sufficient to obtain convergence for the XUV pulses considered in the present work. For the ITP and RTP we use an adaptative Runge-Kutta propagator. The typical time step $\Delta t$ ranges from $10^{-4}-10^{-3}$~atomic units.

\subsection{Ground state}
\label{sec:ground_state}
\begin{table}\centering
\caption{\label{tab:ground_state_energies} Ground state energies of Be for several RAS schemes. $M_1$ and $M_2$ denote the numbers of spatial orbitals in $\mathcal{P}_1$ and $\mathcal{P}_2$ (Fig.~\ref{fig:fig1}), respectively. When all the orbitals are in $\mathcal{P}_1$ ($M_1=M$ and $M_2=0$), the TD-RASSCF approach is equal to the MCTDHF approach with $M=M_1$ orbitals. When, for Be, $M=M_1=2$, the TD-RASSCF approach is equal to the TDHF approach.
The $M=3$ entry denoted by \emph{random} is an example, where we use a random initial wave function. In the other cases the initial guess wave function was designed as described in the text.}
\begin{ruledtabular}
  \begin{tabular}{ccccc}
    Number of orbitals & $M_1$ & $M_2$ &Number of & Ground state\\
    $M=M_1+M_2$ &  &  & configurations & energy (a. u.)\\
    \hline
    2 & 2 & 0 & 1 &-14.57330\\
    \hline
    3 & 3& 0&9& -14.58734 (\emph{random}) \\
      & 3& 0&9&-14.59087 \\
      & 2 &1&5&-14.59087\\
    \hline
    4& 4 & 0&36&-14.60553\\
     & 3 & 1& 18 &-14.60511\\
     & 2 & 2&19 &-14.60551\\
    \hline
    5 & 5& 0&100&-14.61957 \\
    & 3& 2&51&-14.61857\\
    & 2&3&43&-14.61843\\
     & 4&1&52&-14.61826\\
    \hline
    6 & 0& 6& 225&-14.63115\\
    & 5&1&125 & -14.63103\\
    & 3&3& 108 & -14.63020\\
    & 2&4& 77 & -14.62561\\
    \hline
    7 & 7& 0&441&-14.63890\\
    & 6& 1&261&-14.63838\\
    & 5& 2& 220&-14.63827\\
    & 4&3&216&-14.63814\\
    & 2&5&121&-14.63684\\
    \hline
    9 & 9 & 0& 1296&-14.65414\\
     & 2 & 7&239&-14.65082
  \end{tabular}
\end{ruledtabular}
\end{table}

To obtain the ground state we perform an ITP for an appropriate initial guess function~\cite{Hochstuhl2010,Hochstuhl2012}. The choice of the initial guess function is crucial, since there exists an infinite number of non-physical standing wave solutions which are local minima of the TD variational principle~\cite{Bardos2010}. To ensure that the ground state obtained in the ITP is not affected by the selection of the initial guess function we may choose to take an initial random wave function ~\cite{Hochstuhl2010,Hochstuhl2014}. 
The coefficients $c_{\alpha \ell m}^{j}$ of the orbitals [Eq.~(\ref{eq:phip})] are chosen random, and  the amplitudes $C_I$ are taken as $\sqrt{(1+\delta)/N}$, 
where $\delta$ is a random number $ 0 \le \delta < 1$ and $N$ the number of configurations.
This strategy, however, turns out to be problematic for many RAS partitions, including the MCTDHF case, due to the many local minima of the energy located in the manifold~\cite{Bardos2010}. Then a more careful consideration of the design of the initial guess wave function becomes mandatory in order to reach a good approximation, ideally the global minimum, for the ground state~\cite{Helgaker2000}. 

One of the great advantages of the TD-RASSCF method is that we can choose the partition in the active space which induces the most important Slater determinants for a given physical process. In the 3D case this choice becomes very important, since an appropriate set of configurations can facilitate the couplings of the spherical harmonics to the ${}^1\text{S}^e$ ground state. For example, in the case of Be, the main configuration for the ground state is $(1s^22s^2){}^1\text{S}^e$, and an appropriate RAS scheme would be two spatial orbitals in the $\mathcal{P}_1$ space and the rest in $\mathcal{P}_2$. The orbitals in  the $\mathcal{P}_1$ space would be close to $s$-type orbitals, whereas the orbitals in  $\mathcal{P}_2$ would be linear combinations of a set of orbitals which together with the orbitals in $\mathcal{P}_1$ can couple to ${}^1\text{S}^e$. As an example we focus on the case of $M=3$ orbitals, included in Table~\ref{tab:ground_state_energies}, where we show the ground state energy for different RAS schemes. On the one hand, we use a random initial guess function for the MCTDHF approach with $M=3$ orbitals and we obtain the energy $-14.58734$ (the entry in Table~\ref{tab:ground_state_energies} denoted by \emph{random}). 
This value of the MCTDHF energy is in agreement with the energy previously reported in Ref.~\cite{Hochstuhl2014}. 
We note that it is very unlikely that the ITP of a random initial wave functions leads to a lower ground state energy.
This is a consequence of the dominating nature of the $s$ orbitals in the minimization of the energy.
On the other hand we design an initial guess function as follows: i) we choose two orbitals as the $1s$ and $2s$ hydrogenic functions for $Z=4$, ii) we set the amplitude of the Slater determinant which contains them equal to $C_1=0.9$, iii) we choose the 3rd orbital randomly and iv) we set the rest of the coefficients of the configurations to the same non-negative value to obtain a normalized wave function. In this case we obtain the lower $M=3$ MCTDHF energy $-14.59087$~shown in Table~\ref{tab:ground_state_energies}.
We note that this design procedure is somewhat similar to first performing 
a HF calculation with $s$ electrons, and then adding a third orbital to perform a MCSCF calculation; a strategy 
often followed in quantum chemistry multi-configurational self-consistent-field calculations. In a sense, this procedure leaves more freedom to the last orbital to adjust in an optimal manner. Also note that the fact that the energy is lower with the designed initial guess wave function, does not guarantee that the global minimum for the energy is found. The variational principle only allows us to conclude that the ground state obtained with this designed initial guess is more accurate than the one obtained from a completely random initial state.
 
 In the case of TD-RASSCF-D, we impose the most important configurations by setting $M_1=2$ and $M_2=1$ and we obtain again $E=-14.59087$~ now using a random initial guess function, obtained by choosing the coefficients and amplitudes as described in the beginning of this section. The difference between these two approximations to the ground state lies in the orbitals. For the ground state with energy $-14.58734$, the angular part of all the orbitals turns out to be $Y_{00}(\Omega)$, that is, the many-electron wave function corresponds to a multiconfigurational Hartree-Fock (MCHF) using $1s,\,2s~\text{and}~3s$ orbitals. However,  for $E=-14.59087$ and ($M_1=2,\,M_2=1$) we obtain that the main contribution to the angular part of the two orbitals $\phi_1~\text{and}~\phi_2$ is spherically symmetric, while the third orbital $\phi_3$ is a linear combination of $Y_{1,1}(\Omega)$ and $Y_{1,-1}(\Omega)$, such that the expansion coefficients~[Eq.~\eqref{eq:phip}] fulfil $c^3_{\alpha 11}=c^3_{\alpha 1-1}$. This condition ensures that the total magnetic quantum number of the ground state is vanishing, consistent with its ${}^1\text{S}^e$ term. Note that the ground state energies of the RAS $(M_1=3,\,M_2=0)$ and $(M_1=2, M_2=1)$ are the same because these two schemes are equivalent~\cite{Miyagi2014b}. This improvement in the ground state energy for $M=3$ orbitals manifests the importance of the mixing of the orbitals with different values of $m$. If this mixing is not permitted, we would require at least 5 orbitals to improve the HF energy,~\ie, the $1s,\,2s,\,2p_{m=-1},\,2p_{m=0}~\text{and}~2p_{m=+1}$ orbitals, to guarantee the coupling to~${}^1\text{S}^e$. 
Let us remark that the ground state energy for MCTDHF with 5 orbitals is lower than the energy obtained from time-independent multiconfigurational Hartree Fock calculations with 5 orbitals with fixed  $\ell$ and $m$~\cite{Morrison1987}, and that the $M=5$ and $M=9$ results, for which case comparison is possible, are in  good agreement with time-independent MCHF results~\cite{FroeseFisher1993}.

For a given number of orbitals $M$, the RAS scheme $M_1=2$ and $M_2=M-M_1$, may be seen as an improvement to the TDHF solution by adding the possibility for double excitations. The wave function associated with this RAS $(2,M-2)$ is formed by $5M^2-21M+23$ configurations, whereas the number of configurations of the MCTDHF is $\frac{1}{4}(M^2-M)^2$. For example, for $M=9$, the number of configurations required for this RAS is $239$, whereas 
it is 1296 for the MCTDHF approach. The discrepancy in the energy is, however, only $\sim 0.3\%$ (see Table~\ref{tab:ground_state_energies}). 
Because of the reduction in configurations, the results in Table~\ref{tab:ground_state_energies} are obtained with the TD-RASSCF-D approach at a reduced computational cost in terms of CPU and memory compared to  those obtained with the MCTDHF approach. 
The number of operations scales with the sum of two major contributions: The calculation of the two-body operator and the integration of the amplitude equations~\cite{Miyagi2014b}. The numerical effort of the first one is discussed in detail in Sec. II.B. The integration of the amplitude equations scales with the number of configurations, $\text{dim}(\mathcal{V}_\text{RAS})$, as $\mathcal{O}(M^4 \text{dim}(\mathcal{V}_\text{RAS}))$.
We will come back
to the reduction in computational cost in connection with the photoelectron spectra discussed below.

\subsection{Photoelectron spectra}
\label{sec:photoelectron_spectra}

\begin{figure}
  \centering
  \includegraphics[width=.7\linewidth]{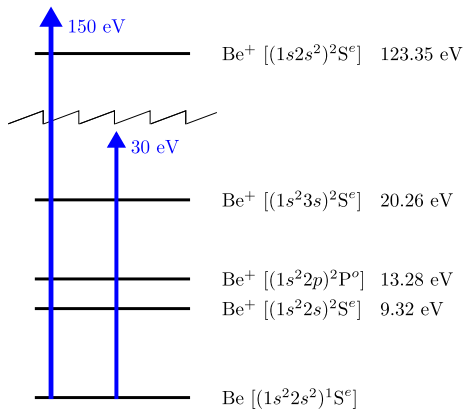}
  \caption{\label{fig:fig2} Energy levels of Be and $\text{Be}^+$ involved in the photoionization process. Experimental energies~\cite{kramida1997} are labelled by their terms and dominant configurations. As indicated by the arrows, we consider lasers with central frequencies corresponding to photon energies of  30~eV and 150~eV. The zig-zag curve above the 30 eV arrow denotes a change in energy scale between the 30 eV and the 150 eV arrows.}
\end{figure}

In this section we illustrate the application of the TD-RASSCF-D method to  photoelectron spectra (PES) of Be after interacting with  short linearly polarized XUV laser pulses. We consider pulses described by the vector potential 
$\vec{A}(t)=A_0\hat{z}\cos^2(\omega t/(2n_p))\sin(\omega t)$, 
where the duration of the pulse is $T=2\pi n_p/\omega$
and the frequency bandwidth $\Delta \omega\approx 1.44\omega/n_p$ with $n_p$ the number of cycles.
The pulse begins at $t=-T/2$. For the photon energies considered in this paper, the photoelectrons come from the ionization of Be to one of the three low-lying states of $\text{Be}^+$: $\text{Be}^+[(1s^22s)^2\text{S}^e]$ (with an ionization potential of $I_{p (1s^22s)^2\text{S}^e}=9.32$~eV), $\text{Be}^+[(1s^22p)^2\text{P}^o]$ ($I_{p (1s^22p)^2\text{P}^o}=13.28$~eV), $\text{Be}^+[(1s^23s)^2\text{S}^e]$ ($I_{p (1s^2 3s)^2\text{S}^e}=20.26$~eV) and 
the higher lying channel $\text{Be}^+[(1s2s^2)^2\text{S}^e]$ ($I_{p (1s2s^2)^2\text{S}^e}=123.35$~eV)~\cite{kramida1997}, see Fig.~\ref{fig:fig2}. 

\subsubsection{30 eV photon energy}
\label{sec:30_ev_photon_energy}

\begin{figure}
  \centering
  \includegraphics[width=.8\linewidth]{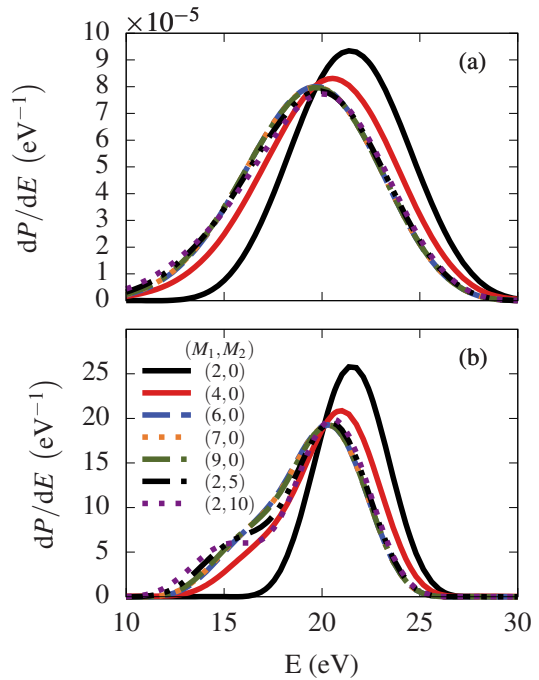}
  \caption{\label{fig:fig3} Photoelectron spectra for  linearly polarized laser pulses with (a) 6 and (b) 10 cycles with a central frequency corresponding to 30~eV, and an intensity of $10^{13}$~W/cm$^{2}$. The RAS schemes are $(M_1,M_2)=(2,0)$~(black, solid), $(4,0)$~(red, solid), $(6,0)$~(blue, long-dashed), $(7,0)$~(orange, dotted), $(9,0)$~(dark green, dash-dotted), $(2,5)$~(black, long dash-double dotted) and $(2,10)$~(dark purple, dotted).} 
\end{figure}

\begin{figure}
  \centering
  \includegraphics[width=.8\linewidth]{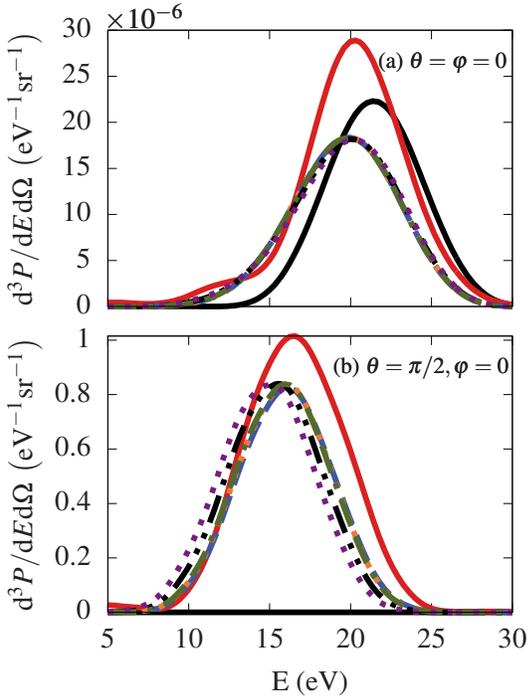}
  \caption{\label{fig:fig4} Triply differential probabilities for $\text{(a)}~\theta=0$ (parallel to the laser polarization direction) $~\text{and}~\text{(b)}~\pi/2$ (perpendicular to the laser polarization direction) and $\varphi=0$ for linearly polarized laser pulses with 6 cycles, central frequency corresponding to  30~eV and an intensity of $10^{13}$~W/cm$^{2}$. The RAS schemes shown are as in Fig.~\ref{fig:fig3}.}
\end{figure}

\begin{figure}
  \centering
  \includegraphics[width=.8\linewidth]{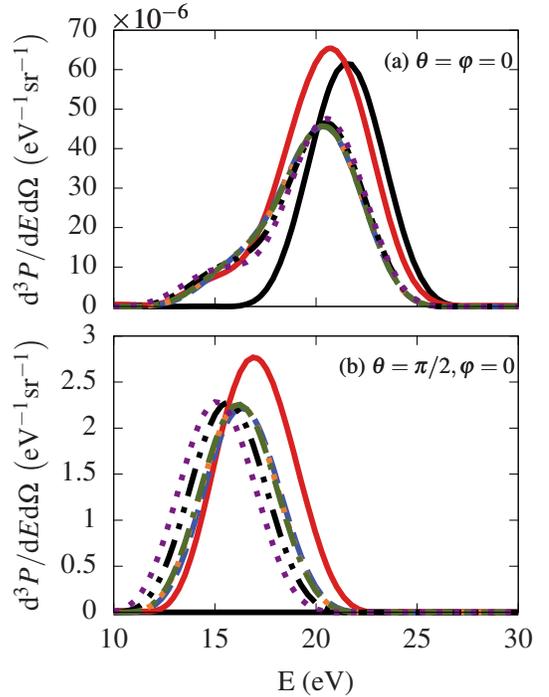}
   \caption{\label{fig:fig5}  Triply differential probabilities for $\text{(a)}~\theta=0~\text{and}~\text{b}~\pi/2$ and $\varphi=0$ for a linearly polarized laser pulse with 10 cycles, a central frequency corresponding to  30~eV and an intensity of $10^{13}$~W/cm$^{2}$. The RAS schemes shown are as in Fig.~\ref{fig:fig3}.}   
\end{figure}

First, we consider  laser pulses with a central frequency corresponding to a photon energy of $30$~eV and an intensity of $10^{13}$~W/cm$^{2}$.  To illustrate the accuracy of the TD-RASSCF-D method, we show the PES after the ionization with pulses containing $n_p=6$ [Fig.~\ref{fig:fig3}(a)] and $n_p=10$ cycles [Fig.~\ref{fig:fig3}(b)] for several RAS schemes. According to the ionization channels in Fig.~\ref{fig:fig2}, the peaks of the photoelectron spectrum should be located at $9.74$, $16.72$ and $20.68$~eV. We first note that the result of the TDHF $(M_1=2,\,M_2=0)$ calculation overestimates  the height of the PES compared with 
the results of the rest of the RAS schemes. This behavior of the TDHF result clearly shows the inadequacy of this approach to describe the electronic structure and dynamics of Be during the ionization process. We find for MCTDHF with 4 orbitals and 6 cycles that the main peak is located at $\approx$ 21.02~eV, and ranges from 20.3-20.5~eV for the rest of the schemes in Fig.~\ref{fig:fig3}(a). The PES for the 10 cycles pulse in  Fig.~\ref{fig:fig3}(b), presents more narrow peaks in the PES, since the bandwidth of the pulse is reduced from $\Delta \omega\sim 7.2$~eV to $4.3$~eV but the peak positions remain similar to the ones in Fig.~\ref{fig:fig3}(a). For the schemes with more than 4 orbitals, the main peak is located at identical positions at the scale of the figure. Also note that the results for MCTDHF for 6, 7 and 9 orbitals are indistinguishable over the entire energy range. In the tail at lower energies, however,  the RAS schemes $(M_1=2,M_2=5)$ and $(M_1=2,M_2 = 10)$ differ from the MCTDHF results. This disagreement lies at energies corresponding to the peak of the ionization channel $\text{Be}~[(1s^22s^2)^1\text{S}^e]\rightarrow\text{Be}^+~[(1s^22p)^2\text{P}^o]$, Fig.~\ref{fig:fig2}. This means that $M_1=2$ and only two electrons in $\mathcal{P}_2$ is not sufficient to describe one electron in an excited orbital and another in the continuum, even for $M_2=10$. To isolate the corresponding peak we use that single ionization changes the angular momentum of the system by $\Delta L=\pm 1$ and $\Delta M_L=0$. Since the ionic channels $(1s^22s)^2\text{S}^e$~and~$(1s^22p)^2\text{P}^o$ have different symmetries, the ejected electron must be $p$ for the first ionic channel and $s~\text{or}~d$ for the second channel. The maximum in the probability of the $p$ continuum electron lies in the direction parallel to the polarization of the laser, whereas it vanishes in the perpendicular direction. The $s$ and $d$ photoelectrons can be detected in both directions. Then, we can disentangle the contributions of the two peaks by analyzing the momentum distributions in the directions parallel and perpendicular to the laser polarizations, such differential quantities are shown in Figs.~\ref{fig:fig4} and~\ref{fig:fig5} for 6 and 10 cycles, respectively. 

In Figs.~\ref{fig:fig4}(a) and \ref{fig:fig4}(b) we show the triply differential probabilities [Eq.~\eqref{eq:triply_diff_energy}] parallel and perpendicular to the polarization of the laser pulse. 
We note that the TDHF is not sufficient to obtain the peak in the perpendicular direction, because the channel Be${}^+[(1s^22p){}^2\text{P}^o]$ is inaccurately described  at that level of approximation.  The reason is that the one-body operator cannot couple directly to that correlated channel.
In Fig.~\ref{fig:fig4}(a) the peak along the polarization direction is smaller than for MCTDHF with 4 orbitals. 
By carefully analyzing the triply differential probability we find that the angular distribution for TDHF is wider which explains the large PES. We see that the probability for ionization in the parallel direction~[Fig.~\ref{fig:fig4}(a)]  is much higher for MCTDHF with 4 orbitals than with the other schemes. Around 10~eV there is a small peak which contributes to the main peak, and which comes from the $s$ and $d$ electrons associated with the Be${}^+[(1s^22p){}^2\text{P}^o]$ channel. The energy and angle resolved signals for the other methods overlap and are indistinguishable on the scale of the figure. For 10 cycles there is, however, a small difference among them between approximately 10 and 17~eV, as shown in Fig.~\ref{fig:fig5}(a). Compared to the 6 cycles case, for 10 cycles, the bandwidth of the pulse is smaller, $\Delta\approx 4.3$~eV, and the influence of the $s$ and $d$ electrons is present in the PES for all the RAS schemes considered. To investigate this difference further,  we turn to the perpendicular direction. For both 6 and 10 cycles, the MCTDHF solution for 6, 7 and 9 orbitals overlap, [Fig.~\ref{fig:fig4}(b) and Fig.~\ref{fig:fig5}(b)], whereas they are different from the RAS schemes $(M_1=2,\,M_2=5)$ and $(M_1=2,M_2=10)$. 
For a 10 cycles pulse, the peaks are located at 16.24 eV for MCTDHF, 15.7 eV for ($M_1=2,M_2=5$) and 15.3 eV for ($M_1=2, M_2=10$). For the 6 cycle pulse, the peaks are shifted by approximately 0.3 eV for all the active spaces. This effect may be caused by the enhancement of other ionization channels opening up due to the wider bandwidth.

\begin{figure}
  \centering
  \includegraphics[width=.8\linewidth]{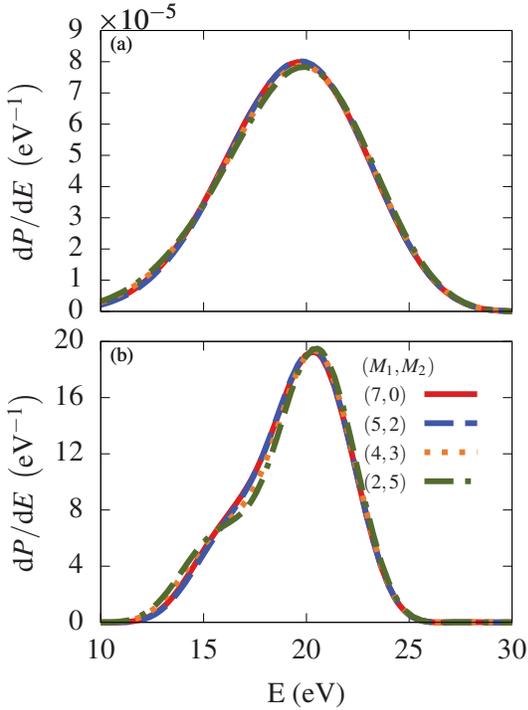}
   \caption{\label{fig:fig6} Photoelectron spectrum for linearly polarized laser pulses with (a) 6 and (b) 10 cycles, a central frequency corresponding to  30~eV,  and an intensity of $10^{13}$~W/cm$^{2}$, for a RAS scheme with 7 orbitals. The RAS schemes are $(M_1,M_2)=(7,0)$~(red, solid), $(5,2)$~(blue, dashed), $(4,3)$~(orange, dotted) and $(2,5)$~(green, dash-dotted)}   
\end{figure}

\begin{figure}
  \centering
  \includegraphics[width=.8\linewidth]{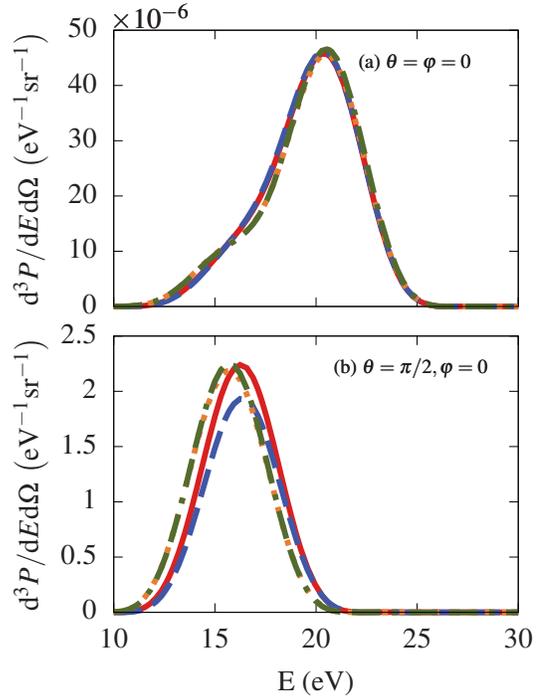}
   \caption{\label{fig:fig7} Triply differential probabilities for $\text{(a)}~\theta=0~\text{and (b)}~\pi/2$ and $\varphi=0$ for a linearly polarized laser pulse with 10 cycles, a central frequency corresponding to 30~eV and an intensity of $10^{13}$~W/cm$^{2}$ for a RAS scheme with $7$~orbitals. The RAS schemes shown are the same are as in Fig.~\ref{fig:fig6}.}   
\end{figure}

To analyze only the influence on the spectra of choosing different RAS schemes, we describe the PES for several partition schemes fixing the number of orbitals. In  Figs.~\ref{fig:fig6} and~\ref{fig:fig7} we show the PES and the triply differential probabilities, respectively, for several RAS schemes for 7 orbitals,~\ie, $(M_1=7,\,M_2=0),(M_1=5,M_2=2),(M_1=4,M_2=3)~\text{and}~(M_1=2,M_2= 5)$. For 10 cycles, the peak at 20.3~eV corresponding to ionization into the ionic $(1s^22s)^2\text{S}^e$ channel as well as the tail for higher energies coincide for all the schemes, whereas for 6 cycles, the peak is shifted to lower energies, probably due to the ionization into the ionic state $(1s^22p)^2\text{P}^o$. As in the previous cases, the disagreement comes at lower energies corresponding to the ionic state $(1s^22p)^2\text{P}^o$. In this range, the MCTDHF and the RAS $(M_1=5,\,M_2=2)$ results can not be distinguished on the scale of Fig.~\ref{fig:fig6}. However, for the RAS $(M_1=4,\,M_2=3)$ there is an excess in the signal for energies lower than 17~eV, whereas for larger energies we find a smaller signal than for MCTDHF. This effect is more pronounced for $(M_1=2,\,M_2=5)$. This can be understood in terms of the differential energy distribution in Fig.~\ref{fig:fig7}. In the parallel direction, the curves for these last two RAS schemes overlap, and they differ between the $(M_1=7,\,M_2=0)$ and $(M_1=5, M_2=2)$ schemes. In the perpendicular direction we can see differences among all the schemes. The peak of the MCTDHF calculation is located at~16.25~eV with a value~$2.2\times 10^{-6}~\text{eV}^{-1}$, which is slightly larger than $1.9\times 10^{-6}~\text{eV}^{-1}$ for $(M_1=5,M_2=2)$. For $(M_1=4,M_2=3)$ and $(M_1=2,M_2= 5)$, the peak is shifted to lower energies, both lying at~15.24~eV. The height of these peaks is very similar to the heights obtained with the MCTDHF results. 

The peak in the PES coming from the ionic state $\text{Be}^+[(1s^23s)^2\text{S}^e]$ is not observed in the present results due to the small cross section of this transition~\cite{Haxton2012}. For the pulses of finite duration used here, this peak is buried in the tails of the other peaks. 

The RTP calculations also illustrate the reduction of the numerical effort of the TD-RASSCF-D approach compared with the MCTDHF method. For the RAS scheme $(M_1=2,M_2=5)$, 110 cycles of propagation  takes 18h in 20 cores  compared to 22h for MCTDHF. In the case of 9 spatial orbitals the difference is more marked, taking 34h for the RAS scheme  $(M_1=2,M_2=7)$ and 47h for MCTDHF.

\subsubsection{150 eV photon energy}
\label{sec:150_ev_photon_energy}

\begin{figure}[b]
  \centering
  \includegraphics[width=.8\linewidth]{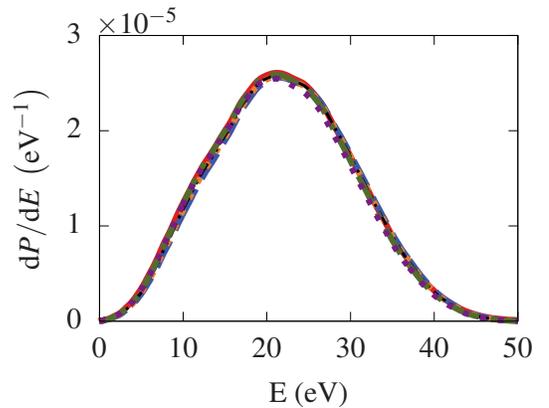}
  \caption{\label{fig:fig8} Photoelectron spectrum for a linearly polarized laser pulse with 10 cycles, a central frequency corresponding to 150~eV  and an intensity of $10^{14}$~W/cm$^{2}$. The RAS schemes shown are as in Fig.~\ref{fig:fig3}.}
\end{figure}

We analyze the core ionization to the ionic channel $\text{Be}^+[(1s2s^2){}^2\text{S}^e]$ [Fig.~\ref{fig:fig2}] using $150$~eV linearly polarized laser pulses for several RAS schemes. The PES is shown in Fig.~\ref{fig:fig8} for a  10 cycles pulse with $\omega=150$~eV and an intensity of $10^{14}$~W/cm$^{2}$. The PES are almost overlapping for all the RAS schemes used, which means that the MCTDHF and TD-RASSCF-D methods describe with the same accuracy the ionization into this channel, even for the minimum number of configurations, that is, using the RAS scheme $(M_1=2,M_2 = M-2)$. These RAS schemes contain the most relevant configurations needed to describe the dynamics of the process, that is, $(1s2s^2)^2$S$^e$ and an electron $\epsilon p$ in the continuum. The peak is located at $\approx$ 21.37~eV, corresponding to an ionization energy of 128.63~eV, a bit higher than the experimental value 123.35~eV~\cite{kramida1997}, which is in agreement with the $4$~eV shift found in Ref.~\cite{Haxton2012}. 

\subsection{Photoelectron dynamics and time-delay}
\label{sec:photoelectron_dynamics_and_time_delay}

\begin{figure}
   \centering
   \includegraphics[width=.95\linewidth]{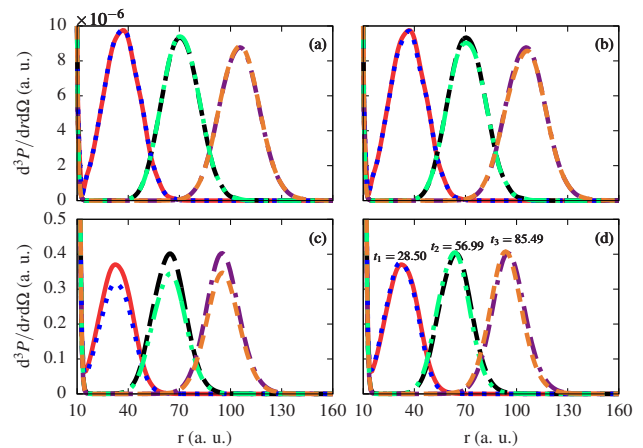}
 \caption{ \label{fig:fig9} Triply differential probabilities in position space for $(M_1=7,\,M_2=0)$,~$(M_1=2,\,M_2=5)$ and~$(M_1=3,\,M_2=4)$ RAS schemes with 7 spatial orbitals after the interaction with a linearly polarized laser pulse with a central frequency corresponding to 30~eV, 10 cycles and an intensity of $10^{13}$~W/cm$^{2}$, where (a) and (b) corresponds to the $\theta=0,\,\varphi=0$ and (c) and (d) to $\theta=\pi/2,\,\varphi=0$. All the panels contain the results for MCTDHF with 7 spatial orbitals at $t_1$ (red, solid), $t_2$ (black, dashed) and $t_3$ (purple, dash-dotted), together with $(M_1=5,M_2=2)$ in panels (a) and (c) ($t_1$: blue, dotted, $t_2$: green, dash-double dotted and $t_3$: orange, short dashed) and $(M_1=4,M_2=3)$ in panels (b) and (d) ($t_1$: blue, dotted, $t_2$: green, dash-double dotted and $t_3$: orange, short dashed). The instants of time in atomic units are indicated in panel (d).}
 \end{figure}

In recent years there has been a large interest in time-delays in photoionization studies (see, e.g., the review Ref.~\cite{Pazourek2015} and references therein). Here we consider the Eisenbud-Wigner-Smith  (EWS) time-delay between ionization into the channels  $\text{Be}^+[(1s^22s){}^2\text{S}^e]$ $(t_{\text{EWS},1s^22s})$ and $\text{Be}^+[(1s^22p){}^2\text{P}^o]$ $(t_{\text{EWS},1s^22p})$ and we exploit the flexibility of the TD-RASSCF method to address the role of electron correlation on photoionization time-delays. We do so in the following by considering the angle-resolved radial density after the pulse for several RAS schemes. We show in Figs.~\ref{fig:fig9}(a)-(b) the density along the polarization direction and in Figs.~\ref{fig:fig9}(c)-(d) the density in the perpendicular direction at different times after the end of the laser pulse. As we have discussed in previous sections, the ionization into both channels contributes in the parallel direction, whereas in the perpendicular direction only the $s$ and $d$ photoelectrons associated with the $\text{Be}^+[(1s^22p){}^2\text{P}^o]$ channel contribute. Let us remark that the results of the MCTDHF calculations have also been obtained with 9 spatial orbitals, and on the scale of the figure they coincide with the results of the calculation for 7 orbitals. In the parallel direction, the height of the density decreases as it spreads during the propagation. In the perpendicular direction the outgoing wavepacket is not yet formed at $t_1=28.50$~a.u., therefore the peak in the density increases for later times. The RAS $(M_1=5,\,M_2=2)$ reproduces accurately the dynamics for ionization into the channel $\text{Be}^+[(1s^22s){}^2\text{S}^e]$ [Fig.~\ref{fig:fig9}(a)], whereas there are small differences with the $(M_1=4,M_2=3)$ scheme, as shown in Fig.~\ref{fig:fig9}(b). The main differences  between the different levels of theory are found in the perpendicular direction shown in Figs.~\ref{fig:fig9}(c) and (d). In the case of $(M_1=5,\,M_2=2)$, we find that the heights of the density peaks are  smaller than in MCTDHF case, but remain on equal positions. In contrast to this case, for $(M_1=4,\,M_2=3)$ the density distributions are similar, but the RAS result is shifted in the radial coordinate with respect to the MCTDHF result. 

To analyze independently the dynamics of the ejected electrons from each ionic channels we can use that the ejected $p$ electron associated with the  ionic 
channel $\text{Be}^+[(1s^22s){}^2\text{S}^e]$ only contributes in the direction parallel to the polarization axis of the laser, whereas the $s$ and $d$ electrons 
associated with the  $\text{Be}^+[(1s^22p){}^2\text{P}^o]$ channel contribute to both the parallel and perpendicular directions. To distinguish between these two 
channels by the angle resolved radial density in the parallel and perpendicular directions we benefit from the fact that the influence of the $s$ and $d$ 
electrons in the parallel direction is negligible compared to the $p$ electron ejected from $\text{Be}^+[(1s^22s){}^2\text{S}^e]$, as seen in Fig.~\ref{fig:fig9}. 
As discussed in Ref. \cite{Pazourek2015} (see also Ref. \cite{Kheifets2010}), the time-delay can be directly extracted from the motion of the outgoing 
wavepacket without the explicit need for  the energy-derivative phase of the dipole matrix element. Accordingly, we may extract the time-delay in the two 
channels by considering the radial density in the parallel and perpendicular directions, i.e., by considering the expectation value of the position of the electron 
in the outer region, which fulfils Ehrenfest's theorem, $\expected{r(t)}=\expected{k}(t-t_0)$, with $t_0$ the 
time-delay, which is a sum of the EWS time-delay and a Coulomb specific contribution due to the logarithmic phase distortion of the outgoing wave 
\cite{Pazourek2015}. The latter is estimated by
$1/\langle k \rangle^3 [1 -\ln (2 \langle k\rangle^2 t)]$, and is a function of $t$ \cite{Pazourek2015}. To isolate the EWS time-delay, the procedure that we will 
follow therefore is to  calculate $t_0$ in the two channels for a finite time interval (in practise the time interval 65-80 a.u. is used), and then subtract the 
contribution from the Coulomb specific time-dependent shift.

We now calculate the time-delays for TD-RASSCF and MCTDHF and compare 
the impact of the electron correlation by varying the RAS scheme. Let us 
remark that we need to include correlation in the photoionization process of Be, 
because we can not resolve the ionization process 
$\text{Be}[(1s^22s^2){}^1\text{P}^e]\rightarrow$ $\text{Be}^+[(1s^22p){}^2\text{P}^o]+e^{-}(s~\text{or}~d)$ 
using TDHF.  We find for MCTDHF with 7 orbitals  a time-delay between the 
two channels of $\tau_\text{EWS}=t_{\text{EWS},1s^22p}-t_{\text{EWS},1s^22s}\approx 
20.81$~as, which is a bit smaller than the result for 9 orbitals, $\tau_\text{EWS}=21.04$~as. 
By reducing $\mathcal{P}_1$ and adding two orbitals to 
$\mathcal{P}_2$, $(M_1=5,\,M_2=2)$, $\tau_\text{EWS}$ is $19.19$~as, which slightly 
increases to $19.4$~as for $(M_1=4,\,M_2=3)$.  The RAS scheme 
$(M_1=2,\,M_2=5)$ leads to $20.19$~as. When we increase the number of orbitals in 
$\mathcal{P}_2$ to $(M_1=2,\,M_2=10)$ we obtain $\tau_0=21.04$~as, 
in agreement with the MCTDHF value for 9 orbitals. 
For these two latter RAS schemes the ionization from the channel $\text{Be}^+[(1s^22p){}^2\text{P}^o]$ 
is not well described, but nevertheless the resulting 
value for $\tau_\text{EWS}$ is acceptable.   The values for the time-delays are collected in Table II.

\begin{table}\centering
\caption{\label{tab:delays} Relative EWS time-delay $\tau_\text{EWS}$ in attoseconds (as) 
between the single photon ionization channels
Be[($1s^2 2s^2)^1$S$^e$] $\rightarrow$ [Be$^+$($1s^22s) + \epsilon  p$] $^1$P$^o$ and
Be[($1s^2 2s^2)^1$S$^e$] $\rightarrow$ [Be$^+$($1s^22p) + \epsilon  \ell$] $^1$P$^o$ with $\ell \in{s,d}$
for 
RAS schemes specified by the values of ($M_1,M_2$). }
\begin{ruledtabular}
  \begin{tabular}{ccc}
        $M_1$ &  $M_2$ &   $ \tau_\text{EWS}$ (as)\\
        \hline
 9 & 0 & 21.04 \\
 7 & 0 & 20.81 \\
 5 & 2 & 19.19 \\
 4 & 3 & 19.40 \\
 2 & 5 &  20.19 \\
 2 & 10 & 21.04
  \end{tabular}
\end{ruledtabular}
\end{table}

In conclusion the relative time-delay of ionization into 
$\text{Be}^+[(1s^22p){}^2\text{P}^o]$ and $\text{Be}^+[(1s^22s){}^2\text{S}^e]$ is around 21 as. 
Note that this method is sensitive to an error in the calculation of $\expected{r(t)}$ and in the 
estimation of the Coulomb shift. It can be estimated that for this process, an 
error of $\Delta\expected{r(t)}=10^{-2}$~a.u. implies an error in the time-delay of the order of 
$\Delta\tau_0\approx 0.8$~as, which together with the error associated to the Coulomb distortion, 
leads to an estimation of the uncertainty in 
$\tau_\text{EWS}$ of around  $\Delta\tau_\text{EWS}\approx 2$~as, 
which is comparable to the experimental accuracy reported in the most recent experiment 
on He (0.9-1.6 as) \cite{Ossiander2017}. 
We also note that the relative accuracy of the different RAS schemes, beyond TDHF,  can not be addressed within the uncertainty 
in the extraction procedure. 

\section{Conclusions and outlook}
\label{sec:conclusions_and_outlook}

In this work we addressed the effect of electron correlation in Be in the ground state, in photoelectron spectra and in relative time-delay in photoionization by application of the TD-RASSCF-D~\cite{Miyagi2013,Miyagi2014b} method, extended in this work to fully 3D systems. We used the coupled basis method on the angular momentum of the single orbital basis to compute the two-body operator. This method reduces the numerical cost since the number of operations to obtain the mean-field operator scales linearly with the radial grid points and quadratically with the number of angular functions. We found that the TD-RASSCF method including double excitations diminishes the numerical effort compared to the MCTDHF by reducing the accessible configurations and that it  is accurate mainly due to the importance of the pairwise nature of the electron correlation. Furthermore, the selection of the RAS makes it possible to identify the most important active space orbitals, which facilitates a convergence to the global ground state. We found that the restriction on the RAS scheme permits a random initial guess function to reach a lower ground state energy than with the MCTDHF method, unless we start with a designed initial guess function in the latter case.

We analyzed the PESs resulting from the interaction with a short linearly polarized XUV laser pulse, and found that the mean-field TDHF method is inaccurate in describing single ionization. Using the TD-RASSCF-D method we identified the most important active space orbitals which capture the most relevant  configurations. 

We also computed an EWS time-delay of around 21~as between the ionization channels  Be${}[(1s^22s^2){}^1\text{S}^e]\rightarrow$ Be${}^+[(1s^22s){}^2\text{S}^e]+e^-$ and Be${}[(1s^22s^2){}^1\text{S}^e]\rightarrow$Be${}^+[(1s^22p){}^2\text{P}^o]+e^-$ for ionization with a few cycle, XUV laser operated in the perturbative regime. We compared the results for several RAS schemes. For example, the TDHF is unable to describe accurately both ionization channels, and therefore, also the time-delay.  As we include more orbitals in the method, we obtain a better description of the ionization into the two channels,
and within the estimated uncertainty of the time-delay, we obtain agreement in the results obtained by different RAS schemes. This agrees with the findings in Ne at different levels of approximation for  photon energies in the range 100-140~eV~\cite{Feist2014a}. 

Over all, we found that the TD-RASSCF-D methodology constitutes an efficient tool to deal with many-electron atomic systems in the presence of time-dependent interaction as, e.g., an external XUV field. Combined with the coupled basis method, it provides a stable, accurate and efficient procedure to treat orbitals with undefined magnetic quantum number, which is crucial to describe, e.g., of the interaction with circularly polarized light, where the rotational symmetry is broken.

\begin{acknowledgments}
We thank Haruhide Miyagi and Camille L\' ev\^ eque for fruitful discussions and comments. This work was supported by the ERC- StG (Project No. 277767-TDMET) and the Villum Kann Rasmussen (VKR) Center of Excellence QUSCOPE. The numerical results presented in this work were obtained at the Centre for Scientific Computing, Aarhus.
\end{acknowledgments}

%

\end{document}